\documentclass[aps,prl,twocolumn,superscriptaddress,showpacs]{revtex4}
\usepackage{graphics,epsfig,amsmath,amssymb,bm}
\begin{document}
\title{Realization of a minimal disturbance quantum measurement}
\author{F. Sciarrino }
\affiliation{Dipartimento di Fisica and Istituto Nazionale per la Fisica della Materia,\\ Universit\`{a} di Roma ''La Sapienza'', p.le A. Moro 5, Roma, I-00185, Italy}
\author{M. Ricci} 
\affiliation{Dipartimento di Fisica and Istituto Nazionale per la Fisica della Materia,\\ Universit\`{a} di Roma ''La Sapienza'', p.le A. Moro 5, Roma, I-00185, Italy}
\author{F. De Martini}
\email[corresponding author:]{francesco.demartini@uniroma1.it}
\affiliation{Dipartimento di Fisica and Istituto Nazionale per la Fisica della Materia,\\ Universit\`{a} di Roma ''La Sapienza'', p.le A. Moro 5, Roma, I-00185, Italy}
\author{R. Filip }
\affiliation{Department of Optics, Palack\'{y} University,\\ 17. listopadu
50, Olomouc 77200, Czech Republic }
\author{L. Mi\v{s}ta, Jr.}
\affiliation{Department of Optics, Palack\'{y} University,\\ 17. listopadu
50, Olomouc 77200, Czech Republic }
\date{\today}

\begin{abstract}
We report the first experimental realization of an ''optimal'' quantum device able to perform a Minimal
Disturbance Measurement (MDM) on polarization encoded qubits
saturating the theoretical boundary established between the
classical knowledge acquired of any input state, i.e. the classical "guess",
and the fidelity of the same state after disturbance due to measurement .
The device has been physically realized by means of a linear optical qubit
manipulation, post-selection measurement and a classical feed-forward
process.
\end{abstract}


\maketitle
The measurement process represents the most innovative and distinctive
aspect of quantum mechanics respect to classical physics. The main result of
the quantum measurement theory is the unavoidable disturbance of the quantum
state induced by the measuring process itself, as epitomized by the early
Heisenberg's X-ray microscope thought experiment \cite{Heis30}. The balance
between the information available on an unknown quantum system and the
perturbation induced by the measurement process is of utmost relevance when
investigating the quantum world \cite{Scul91,Durr98,Bert98,Engl96}. In spite
of this relevance, only in the last years and in the context of Quantum
Information (QI) for finite dimensional systems, an exact quantum
theoretical formulation of this problem has been developed \cite{Bana01}.
When measuring an unknown quantum system $\left| \phi \right\rangle $ two
main questions arise: A: how good is the estimation of the state
obtained by the measuring process? and B: how much the final state is
close to the input one? Adopting the tools developed within QI, the
previous questions can be answered by introducing suitable quantitative
figures of merit to assess the classical information acquired on the state
and the resemblance of the final quantum system to the initial one \cite
{Niel00}. The classical guess $G$ attained by applying a state estimation
strategy is defined as the mean overlap between the unknown state $\left|
\phi \right\rangle $ and the state inferred from the measurement $\rho _{G}$%
: $G=\left\langle \phi \right| \rho _{G}\left| \phi \right\rangle $ while
the closeness of the output quantum state $\rho _{F}$ to the input one is
expressed by the quantum fidelity $F=\left\langle \phi \right| \rho
_{F}\left| \phi \right\rangle $. The final problem is then to establish
which kind of relation connects these two quantities. The higher is the
information achieved the higher is the disturbance applied, and vice versa.
For instance to carry out an optimal state estimation strategy on $\left|
\phi \right\rangle $ we should perform a von Neuman measurement \cite
{Mass95}, i.e. a ''strong disturbance'' one, thus leading to the maximal
modification of the initial state. In this case the output quantum fidelity
is identical to the classical one. This represents an extreme point of the $%
F-G$ boundary. 
\begin{figure}
	\centering
		\includegraphics[height=7cm, width=7cm]{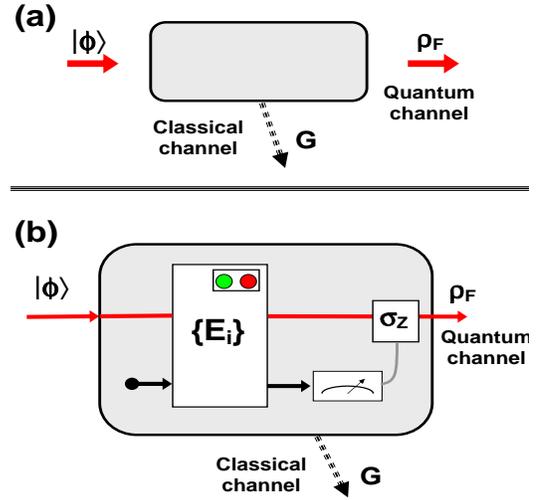}
	\caption{({\bf a}) Diagram of a minimal disturbance measurement ($MDM$)
performed on a single qubit in the state $\left| \phi \right\rangle $.\ The
device provides an output state $\rho _{F}$ with fidelity $F$ and a
''classical guess'' $G$. ({\bf b}) Realization of a $MDM$ by the projector $%
\left\{ E_{i}\right\} $, the measurement of the probe qubit, and the
classical feed-forward $\sigma _{Z}$.}
	\label{fig:MDMFig1}
\end{figure}
On contrary if we want to maintain unchanged the quantum
state, i.e.$F=1$, we can not obtain any previous information about it. This
point defines the other extreme of the $F-G$ plot.

In the present work we consider the basic element of quantum information,
the qubit, which is encoded in a 2-dimensional quantum system and represents the quantum
analogue of the classical bit. Let us start from the situation in which no 
{\it a priori} information is available on the qubit, i.e. this one belongs
to the ''{\it universal}'' set of input states $\left| \phi
_{univ}\right\rangle =\alpha |0\rangle +\beta |1\rangle $ with any $\alpha
,\beta \in C$, $\left| \alpha \right| ^{2}+\left| \beta \right| ^{2}$=%
$1$. The optimal trade-off condition between $G_{univ}$ and $F_{univ}$ was found
by K. Banaszek \cite{Bana01} and reads 
\begin{equation}
F_{univ}\leq \frac{2}{3}+\frac{\sqrt{1-(6G_{univ}-3)^{2}}}{3}  \label{Bana6}
\end{equation}
The two extreme situations outlined above correspond to the points $%
(F_{univ}=2/3$, $G_{univ}=2/3)$ and $(F_{univ}=1$, $G_{univ}=1/2)$. When
partial {\it a priori} information on the qubit to be measured is available,
a better guess of the state can be attained introducing at the same time
less disturbance on the system. Within this framework, a particular simple
case is represented by the set of states called {\it phase} {\it qubits},
for which the information is encoded in the phase $\varphi _{i}$ of the
input qubit represented by any point on any equatorial plane $i$ of the
corresponding Bloch sphere, i.e. $\left| \phi _{cov}\right\rangle \ $= $%
2^{-1/2}(|\Psi _{+}\rangle +e^{i\varphi _{i}}|\Psi _{-}\rangle )$ for a
convenient orthonormal basis \{$|\Psi _{+}\rangle ,|\Psi _{-}\rangle $\}.
For $|\Psi _{\pm }\rangle =2^{-1/2}(|0\rangle \pm i|1\rangle )$ we have $%
\left| \phi _{cov}\right\rangle =\cos \gamma |0\rangle +\sin \gamma
|1\rangle $. {\it Phase} {\it qubits} are adopted in most of the QKD
cryptographic protocols \cite{Gisi02} and the trade-off between phase
estimation and disturbance limited fidelity lies at the basis of the
security assessment problem. In this simpler case the quantum bound reads 
\cite{Mist05}: 
\begin{equation}
F_{cov}\leq \frac{3}{4}+\frac{\sqrt{1-(4G_{cov}-2)^{2}}}{4}  \label{Bana4}
\end{equation}
while the two extreme situations correspond to the points $%
(F_{cov}=3/4,G_{cov}=3/4)$ and $(F_{cov}=1,G_{cov}=1/2)$.

Let us now describe the procedure which saturates the quantum mechanical
bounds, that is, performs the $MDM$ protocol in our work: Fig.1-(b). The
main idea underlying the physical apparatus is to exploit suitable
interaction of the input qubit with an {\it ancilla qubit, }i.e. a{\it \
probe,} and subsequently measure the ancilla to extract information about
the system that we want to guess. By varying the ancilla read-out we are
able to tune the strength of the measurement on the input qubit ranging from
the maximum extraction of \ achievable classical information, i.e. leading
to maximum state disturbance, to no collection of classical information
leaving the input qubit completely unchanged. Let's the ancilla $P$ to be
prepared in the state $2^{-1/2}(|0\rangle _{P}+|1\rangle _{P})$ and the
input qubit $S$ in the generic: $\left| \phi _{univ}\right\rangle
_{S}=\alpha \left| 0\right\rangle _{S}+\beta \left| 1\right\rangle _{S}$.
The interaction between input qubit and probe is achieved by verifying the
''parity'' of the two qubits when they are expressed in the computational
basis ${|0\rangle ,|1\rangle }$. Such {\it parity check }operation entangles
the two qubits when they are in a superposition state of the basis vectors.
To perform this inspection we apply the mutual orthogonal projectors $E_{0}=%
\left[ \left| 0\right\rangle \left| 0\right\rangle \left\langle 0\right|
\left\langle 0\right| +\left| 1\right\rangle \left| 1\right\rangle
\left\langle 1\right| \left\langle 1\right| \right] $, commonly referred as 
{\it parity check} operator \cite{Pitt02}, and $E_{1}=I-E_{0}$ where $I$ is
the identity operator. After successful implementation of the $E_{i}$
projection with probability equal to $1/2$ independently from the input
state $|\phi \rangle $, the overall output state reads $|\Phi
_{i}^{out}\rangle _{SP}=2^{-1/2}(\alpha |0\rangle _{S}|i\rangle _{P}+\beta
|1\rangle _{S}|{i\oplus 1}\rangle _{P})$ where the symbol $\oplus $ denotes
the sum operation modulo 2. Let us consider the case in which $E_{0}$ is
realized. The ancilla is measured in the rotated basis $\left\{
|G_{0}\rangle =\cos \theta |0\rangle +\sin \theta |1\rangle ,|G_{1}\rangle
=\sin \theta |0\rangle -\cos \theta |1\rangle \right\} $. The parameter $%
\theta $ determines the strength of the measurement, The value $\theta =0$
corresponds to optimal state estimation process, i.e maximum $G$ while for $%
\theta =\frac{\pi }{4}$ the input qubit is left unchanged, $F=1$. If the
measurement is successful for the ancillary state $|G_{0}\rangle $ $($or $%
|G_{1}\rangle )$ then the input qubit is guessed to be in the state $\left|
0\right\rangle $ $($or $\left| 1\right\rangle ).$ To complete the protocol,
a unitary operator is applied on the qubit $S$ depending from the
measurement outcome on the probe. In particular if the state $|G_{1}\rangle $
is detected the operation $\sigma _{Z}$ is applied, that is, $\left|
0\right\rangle \rightarrow \left| 0\right\rangle $ and $\left|
1\right\rangle \rightarrow -\left| 1\right\rangle $, while no operation is
applied when the state $|G_{0}\rangle $ has been measured. A similar
procedure is applied when the $E_{1}$ is successful. In this case however
the role of the states $|G_{0}\rangle $, $|G_{1}\rangle $ must be inverted,
that is, $|G_{0}\rangle $ $(|G_{1}\rangle )$ corresponds to $\left|
1\right\rangle $ $(\left| 0\right\rangle )$ and the $\sigma _{Z}$ is
triggered by a click of the $|G_{0}\rangle $ detector. In summary, after the
projection, the measurement of the probe and the feed-forward, the output
qubit density matrix $\rho _{F}$ is achieved by tracing over the probe
Hilbert space and is found in the state: $\rho _{F}=|\phi _{G_{i}}\rangle
\langle \phi _{G_{i}}|+|\phi _{G_{i\oplus 1}}\rangle \langle \phi
_{G_{i\oplus 1}}|$ where $|\phi _{G_{i}}\rangle =\alpha \cos \theta
|0\rangle +\beta \sin \theta |1\rangle $ and $|\phi _{G_{i\oplus 1}}\rangle
=\alpha \sin \theta |0\rangle +\beta \cos \theta |1\rangle $. At the same
time the input state is guessed to be in the state $\rho
_{G}=p_{G_{i}}|0\rangle \langle 0|+p_{G_{i\oplus 1}}|1\rangle \langle 1|$
where $p_{G_{i}}=\left| \alpha \right| ^{2}\cos ^{2}\theta +\left| \beta
\right| ^{2}\sin ^{2}\theta $ and $p_{G_{i\oplus 1}}=\left| \alpha \right|
^{2}\sin ^{2}\theta +\left| \beta \right| ^{2}\cos ^{2}\theta $. From the
previous results we obtain the {\it state-dependent} quantum fidelity and
the classical guess as a function of the parameters {$\alpha ,\beta $}: $%
F_{\phi }=1-2|\alpha |^{2}|\beta |^{2}(1-\sin 2\theta )$ and $G_{\phi
}=\langle \phi |\rho _{G}|\phi \rangle =\frac{1}{2}+\frac{\cos 2\theta }{2}%
(1-4|\alpha |^{2}|\beta |^{2}).$ By averaging the classical guess and the
output fidelity over the ensemble of possible input qubit states, we obtain $%
G_{univ}$=$[3+\sin \left( 2\theta \right) ]/6$ and $F_{univ}$= $[2+\cos
\left( 2\theta \right) ]/3$, which saturate the inequality given by Eq.(\ref
{Bana6}). Interestingly, the previous scheme can also be applied to input 
{\it phase qubit}s states belonging to the equatorial plane of the Bloch
sphere, characterized by real value of the parameters $\left\{ \alpha ,\beta
\right\} $. In this case the average classical guess and output fidelity
are, respectively, $F_{cov}$= $
[3+\sin \left( 2\theta \right) ]/4$ and $G_{cov}=%
[2+\cos \left( 2\theta \right) ]/4$ that satisfy the inequality given by Eq.(%
\ref{Bana4}).
\begin{figure}
	\centering
		\includegraphics[height=7cm]{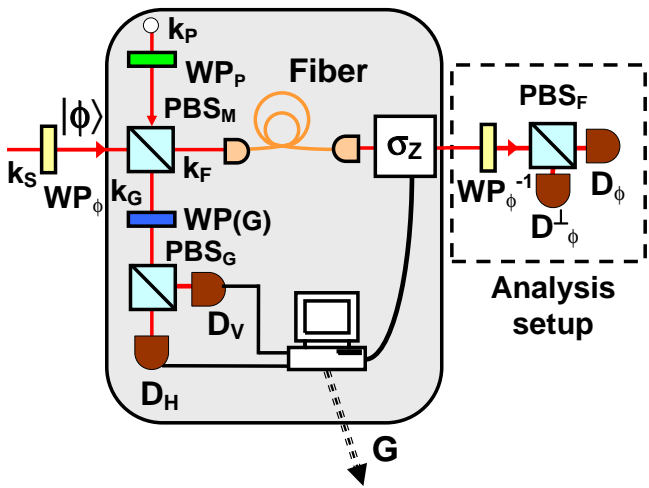}
	\caption{Optical set-up implementing the $MDM$. The output is characterized
adopting the analysis setup illustrated in the dashed box.}
	\label{fig:MDMFig2}
\end{figure}
Let us now turn out attention to the actual implementation of the protocol
for qubits encoded in the polarization state of a single photon by adopting
the isomorphism $|0\rangle \equiv |H\rangle ,|1\rangle \equiv |V\rangle $
where $|H\rangle ,|V\rangle $ denote the horizontal and vertical
polarizations respectively. In order to carry out the projective operations
we have exploited the interference of the two photons, the input qubit to be
measured and the ancilla, at the layer of a polarizing beam splitter $%
(PBS)$, $PBS_{M}$ in Fig.2. $PBS$ transmits the horizontal polarization and
reflects the vertical one, thus when injecting the $PBS$ with a single
photon for each input mode, the successful implementation of the ``parity
check'' $E_{0}$ operator corresponds to the emission of one photon for each
output mode. Indeed this event implies that photons are simultaneously both
transmitted or reflected then exhibiting the same parity. The signature of a
success event is the detection of a single photon in the probe output.
Actually the occurrence of the $E_{0}$ operator was experimentally
identified by detecting a photon on each output mode$.$ Such requirement is
not a limitation since any linear optics quantum information protocols ends
up in photon number measurements of all involved modes \cite{Pryd04}. On the
contrary, the implementation of the projector $E_{1}$ is associated with the
emission of a 2-photon state in one of the two output modes. In this case,
since the two photons are indistinguishable, the signal and the probe qubits
can not be correctly addressed and these events are then discarded. The
experimental $MDM$ device works therefore with probability $p=1/2$.
Nevertheless this probabilistic feature does not spoil the main physical
result of the present procedure since the trade-off conditions are not
altered by any probabilistic procedure \cite{Fiur04,Fili05}.

In the present experiment two photons with equal wavelength (wl) $\lambda
=795nm$ and with a coherence-time $\tau _{coh}=600fs$ were generated in a
non-entangled state on the modes $k_{S}$ and $k_{P}$, Fig.2, by spontaneous
parametric down conversion (SPDC) in a type I BBO crystal in the initial
polarization product state $\left| H\right\rangle _{S}\left| H\right\rangle
_{P}$ \cite{DeMa05}$.$ The input qubit was codified on the mode $k_{S}$ into
the {\it polarization} state $\left| \phi \right\rangle _{S}=\alpha \left|
H\right\rangle _{S}+\beta \left| V\right\rangle _{S}$ by means of a half and
a quarter waveplates $(WP_{\phi })$, whereas the ancilla qubit was
polarization encoded in the state $2^{-1/2}(\left| H\right\rangle
_{P}+\left| V\right\rangle _{P})$ adopting the half-waveplate $WP_{P}$. The
photons $S$ and $P$ were then injected on the two input modes of the
polarizing beam splitter $PBS_{M}$ with an adjustable mutual temporal delay $%
\Delta t$. The condition $\Delta
t=0$ has been identified in a previous experiment to ensure the optimal
temporal overlap of the two photon wavepackets at the PBS layer and hence to
maximize their mutual interference.

The mode $k_{F}$ corresponds to the output quantum channel of the $MDM$
device, while the photon belonging on mode $k_{G}$ enters the classical
measurement apparatus adopted to infer the classical guess $G$. This
estimation task is realized by means of a tunable half-waveplate $WP(G)$, a
polarizing beam splitter $PBS_{G}$, and two detectors $D_{H}$, $D_{V}$ . The
angular position of $WP(G)$, $\vartheta _{G}=\theta /2$, determines the
strength of the measurement. The complete protocol implies a classical
feed-forward on the polarization state of the photon belonging to the mode $%
k_{S}$ depending on which detector $(D_{H}$ or $D_{V})$ is fired: precisely
if the detector $D_{V}$ clicks, a $\sigma _{Z}$ Pauli operation is applied,
in the other case no transformation is implemented on the quantum channel.
To carry out the $\sigma _{Z}$ transformation we adopted a fast $LiNbO_{3}$
Pockels cell (PC) electronically driven by a transistor array activated by a
click of detector $D_{V}$. The $\sigma _{Z}$ transformation was implemented
by a applying to the PC a $\lambda /2$ voltage, i.e., leading to a $\lambda
/2$ induced phase shift of the $\left| V\right\rangle $ polarization
component. Details on the electronic circuit piloting the electro-optic
Pockels cell can be found in Ref. \cite{Giac02}. In order to synchronize the
active window of the Pockels cell with the output qubit, the photon over the
mode $k_{F}$ was delayed through propagation over a $30m$ long single mode
optical fiber. The polarization state on the mode $k_{F}$ after the
propagation through the system fiber+PC was analyzed by the combination of
the waveplate $WP_{\phi }^{-1}$ and of the polarization beam splitter $%
PBS_{F}$. For each input polarization state $\left| \phi \right\rangle _{S}$%
, $WP_{\phi }^{-1}$ was set in order to make $PBS_{F}$ to transmit $\left|
\phi \right\rangle $ and reflect $| \phi ^{\perp }\rangle $.
\begin{figure}
	\centering
		\includegraphics[height=9cm,width=7cm]{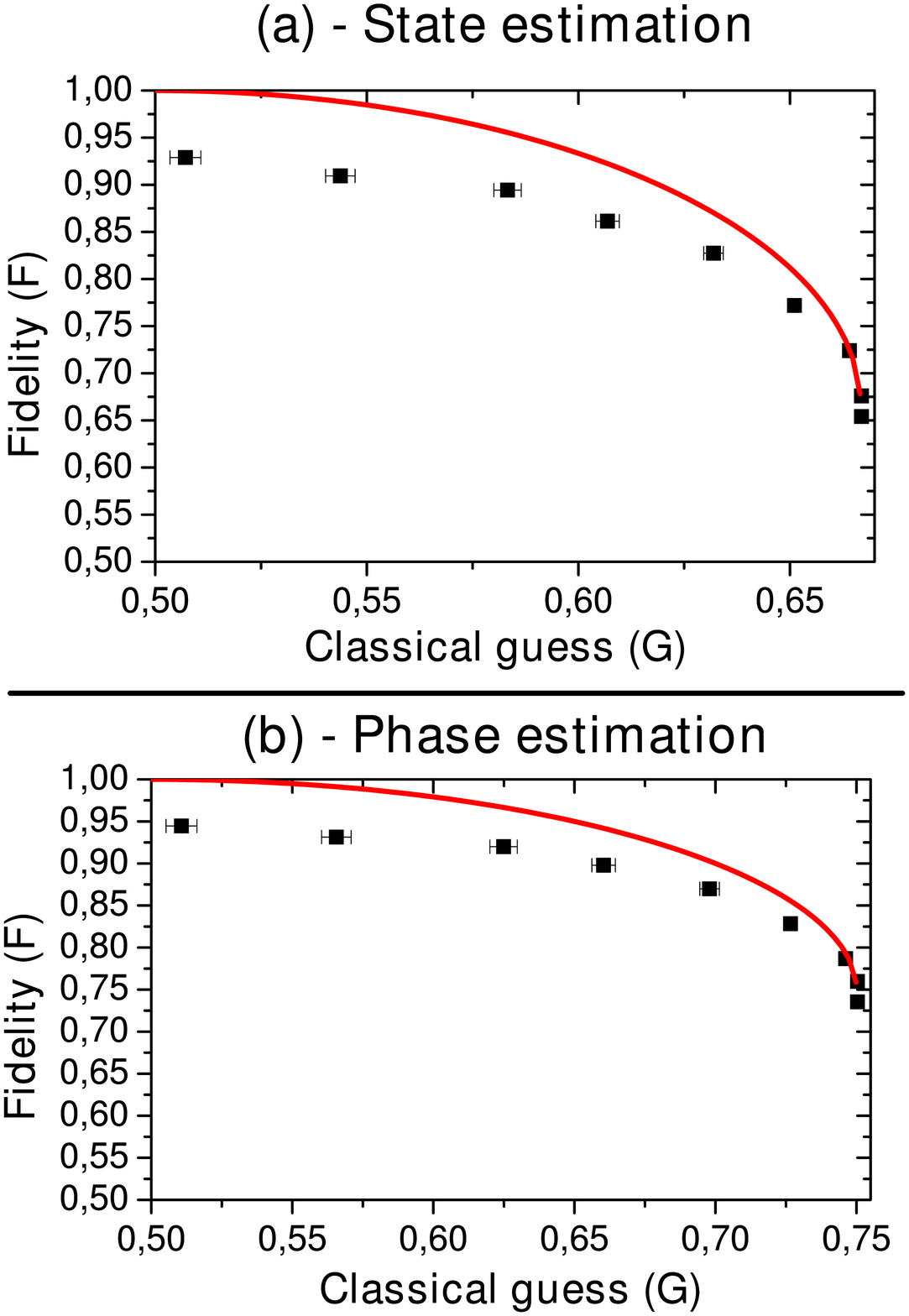}
	\caption{({\bf a}) Experimental data of the quantum fidelity $F$ versus the
classical guess $G$ for an arbitrary input qubit. The fidelities have been
averaged over the six states $\left\{ \left| H\right\rangle ,\left|
V\right\rangle ,\left| L_{\pm }\right\rangle ,\left| C_{\pm }\right\rangle
\right\} $; solid line: optimal trade-off between $F_{univ}$ and $G_{univ}$
(Eq.1); ({\bf b})\ Experimental data of the quantum fidelity $F$ versus the
classical guess $G$ for an equatorial input qubit. The fidelities have been
averaged over the four states $\left\{ \left| H\right\rangle ,\left|
V\right\rangle ,\left| L_{\pm }\right\rangle \right\} $; solid line: optimal
trade-off between $F_{cov}$ and $G_{cov}$ (Eq.2).}
	\label{fig:MDMFig3}
\end{figure}
Two different experiments have been carried out. In the first one the device
has been characterized either for a universal set and for a covariant set of
input qubits. To demonstrate the realization of the $MDM$ apparatus it is
sufficient to use a finite set of non-orthogonal quantum states from
mutually maximally-complementary bases. For the universal $MDM$, we have
adopted the three maximally complementary basis {$|H\rangle $, $|V\rangle $}%
, {$|L_{\pm }\rangle =2^{-1/2}(|H\rangle \pm |V\rangle )$} and {$|C_{\pm
}\rangle =2^{-1/2}(|H\rangle \pm i|V\rangle )$} whereas for phase
covariant $MDM$ we employed the {$|H\rangle $, $|V\rangle $} and {$|L_{\pm
}\rangle $} basis only. Such sets of states are adopted in the conventional
quantum cryptographic protocols \cite{Gisi02}. For each state $|\phi \rangle
_{S}$, the corresponding values of $F_{\phi }$ and $G_{\phi }$ were measured
for different $\vartheta _{G}$ settings. This task was achieved by
collecting the 2-fold coincidences between the two sets of detectors $%
\left\{ D_{H},D_{V}\right\} $\ and $\{ D_{\phi },D_{\phi }^{\perp
}\} $ and then extracting the joint probabilities of the two-photon
states $p_{H\phi },$ $p_{V\phi },$ $p_{H\phi \perp },$ $p_{V\phi \perp }$
where $p_{ij}$ is the relative frequency of the coincidence counts $%
D_{i}-D_{j}.$ The fidelity of the output state $\rho _{out}$ can be
evaluated as $F_{\phi }=\langle \phi |\rho _{out}|\phi \rangle =p_{H\phi
}+p_{V\phi }.$ To extract the value $G_{\phi }$, we first calculate the
occurrence probability $P_{i}$ $\{i=H,V\}$ of the measurement $|i\rangle
\langle i|$, as $P_{i}=p_{i\phi }+p_{i\phi \perp }.$ In this case the input
state is guessed to be in the quantum state $|i\rangle $ leading to a
fidelity $|\langle \phi |i\rangle |^{2}.$ Hence for each state $|\phi
\rangle $ the resulting estimation fidelity is obtained as $G_{\phi
}=\sum_{i}P_{i}|\langle \phi |i\rangle |^{2}.$ The mean quantum fidelities
and classical guesses were averaged over all the input states. The
experimental data are reported in Fig.3. For the ``universal'' $MDM$ the
extreme experimental points are $(G_{univ}^{exp}=0.666\pm
0.001;F_{univ}^{exp}=0.654\pm 0.004)$ and $(0.507\pm 0.004;$ $0.929\pm
0.002),$ corresponding to the settings $\vartheta _{G}=0^{\circ }$ and $%
\vartheta _{G}=22.5^{\circ }.$ These figures are to be compared with the
theoretical limits: $(G_{univ}^{th}=0.666$; $F_{univ}^{th}=0.666)$ and $(0.5$%
; $1)$. Likewise, for the ``phase covariant'' $MDM$ the extremal
experimental points are $(G_{cov}^{exp}=0.750\pm 0.001$; $%
F_{cov}^{exp}=0.735\pm 0.004)$ and $(0.511\pm 0.006;$ $0.945\pm 0.003)$ to
be compared with the theoretical: $(G_{univ}^{th}=0.75$; $%
F_{univ}^{th}=0.75) $ and $(G_{univ}^{th}=0.5$; $F_{univ}^{th}=1)$. The
deviations from the theoretical curves are mainly due to the $PBS$
imperfections $\left( R_{H}\simeq 3\%\right) $ which partially spoil the
''parity check'' operation..

In summary, we realized conditional implementation of Minimal Disturbance
Measurement saturating the quantum mechanical $F-G$ trade-off, both for
universal and for phase covariant set of states. The present
procedure can be adopted for different qubit hardware and can have
interesting applications in the framework of quantum communication to improve the transmission fidelity of a lossy quantum channel\cite
{Ricc05}.

We thank J. Fiur\'a\v{s}ek for fruitful discussions. F.D.M., M.R., and F.S. acknowledge financial
support from the Ministero della Istruzione, dell'Universit\'{a} e della
Ricerca (COFIN 2002). R.F. and L.M. acknowledge
support from Research project Measurement and information in Optics of Czech
Ministry of Education. R. F. thanks to support by Project 202/03/D239 of the
Grant Agency of Czech Republic.

\end{document}